# A method to restore the intrinsic dielectric functions of 2D materials in periodic calculations and its applications to the dielectric and optical properties of ultrathin h-BN and MoS$_2$

*Guang Yang, Shang-Peng Gao\**

Department of Materials Science, Fudan University, Shanghai 200433, P. R. China

Previous calculations of the dielectric and optical properties of 2D materials often overlooked or circumvented the influence of vacuum spacing introduced in periodic calculations, which gave rise to mispredictions of the intrinsic properties of 2D materials or merely qualitative results. We first elucidate the relationship between the vacuum spacing and the dielectric and optical properties of 2D materials in periodic calculations, and then bring forward an effective method to accurately predict the dielectric and optical properties of 2D materials by restoring the intrinsic dielectric functions of 2D materials independent of the additional vacuum spacing. As examples, the intrinsic dielectric and optical properties of ultrathin h-BN and MoS$_2$ from monolayer to pentalayer, including dielectric functions, optical absorption coefficients, refraction indexes, reflectivities, extinction coefficients, and energy loss functions, have been calculated by our method. Our calculations reveal that the out-of-plane optical dielectric constants, static refraction indexes, and static reflectivities of 2D h-BN and MoS$_2$ increase as the number of layers increases, while the in-plane counterparts remain unchanged. Excitonic frequency-dependent optical properties of h-BN and MoS$_2$ from monolayer to bulk are also calculated by solving the Bethe-Salpeter equation and show strong anisotropy. In better agreement with experimental results than previous calculations, the presented method demonstrates enormous potential to investigate the dielectric and optical properties of other 2D materials extensively and quantitively.

# I. INTRODUCTION

Atomically thin two-dimensional (2D) materials have gathered increasing attention in multiple research areas, including catalysis, electronics, and photonics [1–5] since the technical obstacles to the synthesis of monolayer or few-layer materials such as graphene [6], hexagonal boron nitride (h-BN) [7], transition metal dichalcogenides (TMDs) [8–10] have continued to be removed. Furthermore, with the constant discovery of new 2D materials including emergent 2D heterostructures [1,11–13], numerous research effort has been directed toward exploring and designing 2D materials with tailored and superior properties in the recent past both theoretically and experimentally [14–25].

So far, plenty of theoretical researches have made predictions on the various properties of 2D materials. They are mainly based on the supercell approach under periodic boundary conditions, which is adopted in calculations with the density functional theory (DFT), many-body perturbation theory (MBPT), quantum Monte Carlo method, and the like. The supercell calculations are conducted by inserting a sufficiently large vacuum layer between the adjacent layers of 2D materials to form the supercell, which can avoid the spurious interaction between 2D materials in neighboring supercells. The ground-state properties of 2D materials, such as structural constants, adsorption properties, and electronic properties, converge rapidly with increasing vacuum spacing. To overcome the relatively slow convergence of the excited properties of 2D materials compared to the ground-state electronic structures, several Coulomb truncation approaches have been developed [26–29]. Unfortunately, even when the convergence of quasiparticle band gap and exciton binding energy is reached in supercell calculations, the dielectric and optical properties of 2D materials, which depend on supercell volume, can still vary with vacuum spacing and converge to the vacuum limit at infinite vacuum spacing, making it difficult to derive truly reliable results. We shall refer to the dielectric and optical properties of the supercell containing the vacuum layer and 2D materials as non-intrinsic values, in comparison with the intrinsic values of the 2D materials independent of the vacuum spacing. As a result, previously calculated non-intrinsic dielectric and optical properties of a specific 2D material cannot be compared directly with each other since different vacuum volumes were used. Even worse, the non-intrinsic dielectric and optical properties of 2D materials deviate significantly from experimental measurements. Recently, the quasi-2D descriptions of screening and polarizability are under development by adapting the calculation algorithms for single-layer and interlayer responses [30–34].

Several previous researches specifically explored the influence of vacuum spacing on the calculated dielectric constants of 2D materials [35–37]. For instance, Weng et al. [35] found the relationship between the vacuum thickness and the static dielectric constant $\varepsilon_0$ from the microscopic perspective of dielectric polarization, and Laturia et al. [36] obtained the dependence of $\varepsilon_0$ and the optical dielectric constant $\varepsilon_\infty$ of 2D materials on the vacuum thickness within the theory of equivalent capacitance of series and parallel combination. These two studies gave the same and reasonable analysis on the static dielectric constant $\varepsilon_0$. However, it is worth noting that $\varepsilon_0$ consists of the electronic contribution $\varepsilon_\infty$ (also known as the optical dielectric constant) and the ionic contribution $ε_l$, and the relationship between the vacuum thickness and $ε_0$ of 2D materials in the work of Laturia et al. cannot apply to $ε_∞$ since the macroscopic charge-discharge process is only suitable for treating $ε_0$. Overall, an extensive study on the role of the vacuum spacing in periodic calculations of the dielectric and optical properties of 2D materials has rarely been reported to our knowledge. Most of the previous researches on the dielectric and optical properties of 2D materials presented directly the non-intrinsic values including the influence of vacuum spacing [17,18,20,38–42], or chose to circumvent the problem by focusing on the relative shape of the imaginary dielectric function of 2D materials [29,43–46], which left the dielectric and optical properties of 2D materials including the real part of dielectric functions, optical absorption coefficients, refraction indexes, reflectivities, extinction coefficients, and energy loss functions uninvestigated.

In this work, we shed light on the relationship between the volume of vacuum region and the in-plane and out-of-plane dielectric functions of 2D materials in periodic calculations by both practical calculations and a close examination of the algorithm of the dielectric functions. Based on this relationship, we formulate a method to restore the intrinsic dielectric functions of 2D materials independent of vacuum spacing from the periodic *ab initio* supercell calculations. As a result, intrinsic optical dielectric constant $ε_∞$ and other intrinsic dielectric and optical properties can be further obtained. Moreover, our restoration method is highly feasible for all dielectric and optical calculations including those at the independent-particle approximation (IPA) or Bethe-Salpeter equation (BSE) level because it only involves a post-treatment of the dielectric functions obtained from supercell calculations without the need to reformulate the computational algorithm and thus avoids sophisticated code editing. As examples, the intrinsic dielectric and optical properties of h-BN and $MoS_2$ from monolayer to pentalayer have been calculated in both in-plane and out-of-plane directions by the BSE method. Strong anisotropy has been found in their dielectric and optical properties and in the dependence of the dielectric and optical properties on the thicknesses of 2D materials. Overall, the out-of-plane properties, including the optical dielectric constants, static refraction indexes, and static reflectivities of h-BN and $MoS_2$, increase as the number of layers is increased from monolayer to bulk, while the in-plane counterparts remain almost unchanged.

Most significantly, the restored intrinsic dielectric and optical properties including the real and imaginary dielectric functions, extinction coefficient, and refraction index of 2D $MoS_2$ agree better with experimental measurements than previous results obtained directly from supercell calculations. Therefore, our method tackles the discrepancy between predictions and measurements and therefore underscores the importance of restoring the intrinsic dielectric and optical properties in periodic calculations. We believe that our method has enormous potential to extensively and quantitatively predict the intrinsic dielectric and optical properties of various 2D materials. In addition, we also put forward an effective method to reasonably compare and evaluate the results of 2D material arrays with different volumes of vacuum region. We hope that the restoration method and the predictions of the intrinsic dielectric and optical properties of 2D h-BN and MoS$_2$ in this work can provide a solid foundation for tailoring and designing the dielectric and optical applications of 2D materials in the future.

## II. RESTORATION METHOD OF THE DIELECTRIC FUNCTIONS OF 2D MATERIALS IN PERIODIC CALCULATIONS

In order to calculate the dielectric and optical properties of 2D materials using the periodic boundary condition, a large vacuum space is often added to isolate the interaction between periodic 2D slabs to form the supercell. Properties like in-plane lattice constants, electronic band structures, and exciton wavefunctions will definitely converge well if the vacuum layer is large enough, but this is not the case for dielectric response functions. In fact, the real and imaginary parts of the dielectric functions have different dependence on the volume of the supercell containing the vacuum layer. Thus, the optical properties obtained via both the real and imaginary dielectric functions can be dramatically altered with the choice of vacuum volume. Now we will reveal this relationship between the dielectric function and vacuum spacing in supercell via volumes of the supercell and 2D material.

We first look at the macroscopic dielectric function $\varepsilon_M$ calculated by the many-body BSE method with the Tamm-Dancoff approximation (TDA) [47]:

$$\varepsilon_M(\omega) = 1 + \lim_{q \to 0} v(q) \sum_\lambda \frac{\left|\sum_{v,c;k} \langle v, k | e^{-iq \cdot r} | c, k \rangle A_\lambda^{(v,c;k)}\right|^2}{E_\lambda - (\omega + i\eta)}$$

$$= 1 + \lim_{q \to 0} v(q) \sum_\lambda [P(E_\lambda - \omega)^{-1} - i\pi\delta(E_\lambda - \omega)] \left|\sum_{v,c;k} \langle v, k | e^{-iq \cdot r} | c, k \rangle A_\lambda^{(v,c;k)}\right|^2 \quad (1)$$

Here in Eq. (1), we have used the relation $(E_\lambda - \omega + i\eta)^{-1} = P(E_\lambda - \omega) - i\pi\delta(E_\lambda - \omega)$, and $v$ and $c$ are short for valence and conduction states. $P$ stands for principal value of the integral, and the matrix element of the Coulomb interaction $v(q)$ can be evaluated in reciprocal space as:

$$v(q) = \frac{1}{\Omega}\sum_G \langle c, k|e^{iG\cdot r}|v, k\rangle v(q+G) \langle v', k'|e^{-iG\cdot r}|c', k'\rangle = \frac{1}{\Omega} M_v \qquad (2)$$

where $\Omega$ is the unit cell volume of the calculated system in real space, and we use $M_v$ to stand for the matrix elements summation part, which does not explicitly contain the volume term.

According to Eqs. (1) and (2), the imaginary dielectric function should be inversely proportional to the volume of the supercell:

$$Im(\varepsilon_M) = \frac{-\pi}{\Omega}\lim_{q \to 0} \sum_\lambda [\delta(E_\lambda - \omega)] \left|\sum_{v,c;k}\langle v, k|e^{-iq\cdot r}|c, k\rangle A_\lambda^{(v,c;k)}\right|^2 \qquad (3)$$

Therefore, in periodic calculations for 2D materials, the absolute scale of imaginary dielectric function $Im(\varepsilon_M)$ should always decrease with the increasing vacuum spacing between 2D slabs, although their spectra characteristics such as relative peak shape should converge when the vacuum spacing is large enough. In this way, the imaginary dielectric function in the supercell calculation with the supercell volume $\Omega_2$ is related to another imaginary dielectric function in the supercell calculation with the supercell volume $\Omega_1$ through Eq. (4), given that both of them are for the same 2D material and the vacuum spacing employed in each calculation is large enough for the electronic structures to converge:

$$Im(\varepsilon_M)_{\Omega 2} = \frac{\Omega_1}{\Omega_2} Im(\varepsilon_M)_{\Omega 1} \tag{4}$$

Hence, for actual 2D materials without vacuum layer, the intrinsic imaginary dielectric function $Im(\varepsilon_M)_{2D}$ in in-plane or out-of-plane directions should be obtained by renormalizing the non-intrinsic imaginary dielectric function of the supercell $Im(\varepsilon_M)_{SC}$ as follows:

$$Im(\varepsilon_M)_{2D} = \frac{\Omega_{SC}}{\Omega_{2D}} \times Im(\varepsilon_M)_{SC} = \frac{d_{SC}}{d_{2D}} \times Im(\varepsilon_M)_{SC} \tag{5}$$

where $d_{SC}$ is the thickness of 2D material $d_{2D}$ plus the thickness of the vacuum layer in the supercell as depicted in Fig. 1.

Meanwhile, the real dielectric function can always be obtained by the Kramers-Kronig transformation from the imaginary part as:

$$Re(\varepsilon_M) = 1 + \frac{2}{\pi} P \int_0^\infty \frac{Im\ (\varepsilon_M)\omega'}{\omega'^2 - \omega^2} d\omega' \tag{6}$$

We can also explicitly write out the volume dependence of the real dielectric function from Eqs. (1) and (2) or via the Kramers-Kronig transformation from imaginary dielectric function in Eq. (3):

$$Re(\varepsilon_M) = 1 + \frac{1}{\Omega} \lim_{q \to 0} M_v \sum_\lambda [P(E_\lambda - \omega)^{-1}] \left| \sum_{v,c;k} \langle v, k | e^{-iq \cdot r} | c, k \rangle A_\lambda^{(v,c;k)} \right|^2 \tag{7}$$

Likewise, for the same 2D materials the real dielectric function in the supercell calculation with the supercell volume $\Omega_2$ is related to another real dielectric function in the supercell calculation with the supercell volume $\Omega_1$ by:

$$Re(\varepsilon_M)_{\Omega 2} = 1 + \frac{\Omega_1}{\Omega_2} [Re(\varepsilon_M)_{\Omega 1} - 1] \tag{8}$$

Vacuum thicknesses should also be large enough for the electronic structures of both supercells to converge. Equation (8) indicates that $Re(\varepsilon_M) - 1$ should decrease with the increasing thickness of the vacuum layer. Therefore, for actual 2D material without vacuum layer the intrinsic real dielectric function $Re(\varepsilon_M)_{2D}$ in in-plane or out-of-plane direction should be reconstructed as the following equation:

$$Re(\varepsilon_M)_{2D} = 1 + \frac{\Omega_{SC}}{\Omega_{2D}} \times [Re\ (\varepsilon_M)_{SC} - 1] = 1 + \frac{d_{SC}}{d_{2D}} \times [Re\ (\varepsilon_M)_{SC} - 1] \qquad (9)$$

It is worth noting that the relations indicated in Eqs. (4), (5), (8), and (9) also hold good for the dielectric functions evaluated using the simpler IPA:

$$\varepsilon_M(\omega) = 1 + \frac{1}{\Omega}\lim_{q \to 0} \frac{4\pi e^2}{q^2} \sum_{nm,k} \frac{2(f_{mk'} - f_{nk})|\langle m, \mathbf{k'}|e^{-i\mathbf{q}\cdot\mathbf{r}}|n, \mathbf{k}\rangle|^2}{(E_{n,k} - E_{m,k'}) - (\omega + i\delta)} \qquad (10)$$

where $f_{mk'}$ and $f_{nk}$ are the occupation numbers of the single-particle electronic states $|m,\mathbf{k'}\rangle$ and $|n,\mathbf{k}\rangle$. Therefore, we believe that the analysis and discussion for the issue of vacuum spacing in periodic calculations of dielectric functions are applicable to 2D materials regardless of the theoretical level: from the simple IPA methods to the advanced many-body approaches.

Above theoretical relations have also been proven to be valid by practical calculations as depicted in Fig. 2: the relationship between the thicknesses of supercells and the dielectric functions of monolayer h-BN in both in-plane and out-of-plane directions calculated by the BSE methods shows great consistency with Eqs. (4) and (8). To sum up, our restoration scheme can transform the non-intrinsic dielectric function of the supercell containing both the 2D material and the vacuum slab to the intrinsic value of the 2D material. Note that the Kramers-Kronig relationship still holds true for dielectric functions, either intrinsic or non-intrinsic. Other intrinsic optical properties including optical absorption α(ω), refraction index n(ω), reflectivity R(ω), energy loss function L(ω), and extinction coefficient k(ω) can be obtained through the real and imaginary macroscopic dielectric functions $Re(\varepsilon_M)$ and $Im(\varepsilon_M)$ as follows:

$$\alpha(\omega) = \omega\sqrt{2\sqrt{Re\ (\varepsilon_M)^2 + Im\ (\varepsilon_M)^2} - 2Re\ (\varepsilon_M)} \quad (11.a)$$

$$R(\omega) = \left|\frac{\sqrt{\varepsilon_M}-1}{\sqrt{\varepsilon_M}+1}\right|^2 \quad (11.b)$$

$$n(\omega) = \sqrt{\frac{1}{2}[\sqrt{Re\ (\varepsilon_M)^2 + Im\ (\varepsilon_M)^2} + Re\ (\varepsilon_M)]} \quad (11.c)$$

$$k(\omega) = \sqrt{\frac{1}{2}[\sqrt{Re\ (\varepsilon_M)^2 + Im\ (\varepsilon_M)^2} - Re\ (\varepsilon_M)]} \quad (11.d)$$

$$L(\omega) = \frac{Im\ (\varepsilon_M)}{Re\ (\varepsilon_M)^2 + Im\ (\varepsilon_M)^2} \quad (11.e)$$

It is worth noting that in our work the thicknesses of n-layer 2D materials $d_{2D}$ are estimated by the distance between outermost atomic centers over and under the plane of (n+1)-layer 2D materials as depicted in Fig. 1(c). There are actually other estimation methods of the thicknesses $d_{2D}$ of n-layer 2D materials: one is the distance $d_I$ between outermost atomic centers over and under the plane of n-layer 2D materials plus the radii $r$ of the outermost ions [see Fig. 1(b)], and the other is the interlayer distance $d_{II}$ in the corresponding bulk materials multiplied by n [see Fig. 1(d)]. We must notice that although different estimation methods will bring slight uncertainty to the dielectric functions compared with the measured thicknesses of 2D materials, the restoration method we introduce in this work still holds true and can be applied to any 2D systems with available thicknesses from estimation or measurement. Hence, as long as we know the explicit volumes of the supercells and 2D materials, their non-intrinsic dielectric and optical properties can always be compared and evaluated based on Eqs. (4) and (8) and transformed to the intrinsic values based on Eqs. (5) and (9).

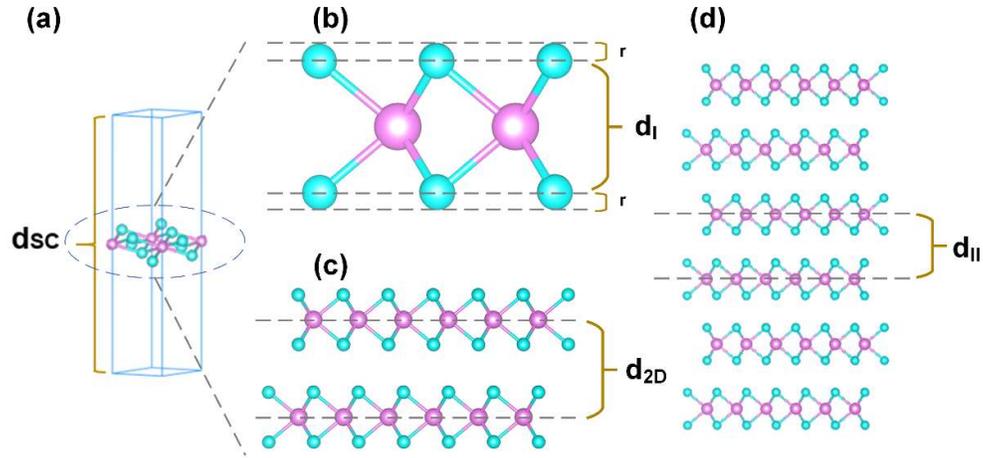

FIG. 1. (a) The supercell model of the 2D material with vacuum spacing, where $d_{SC}$ is the thickness of 2D material plus the vacuum spacing in the supercell. (b) The monolayer and (c) bilayer model of the 2D material without vacuum spacing. (d) The model of the bulk material comprised of layers of 2D material. Note that the thickness of the monolayer material can be obtained as $d_I + 2r$ in (b), $d_{2D}$ in (c), or $d_{II}$ in (d).

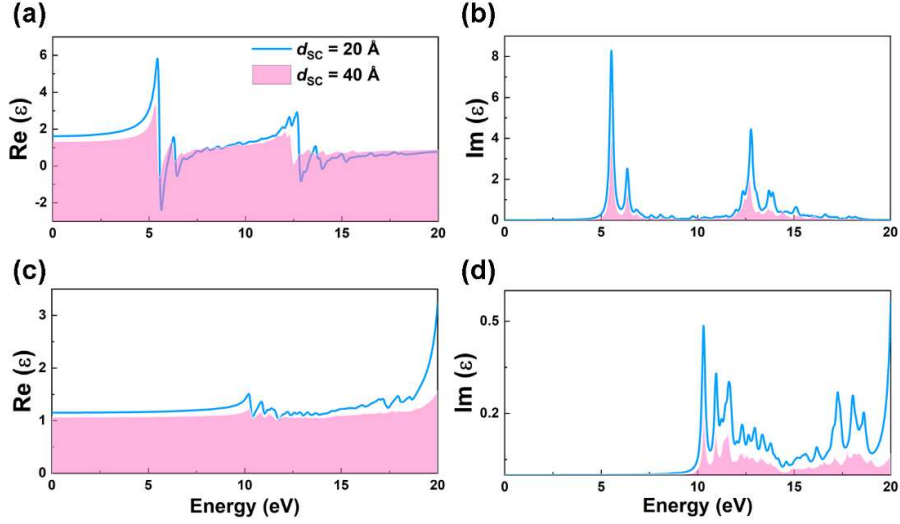

FIG. 2. (a) Real and (b) imaginary part of the dielectric response function of monolayer h-BN in the in-plane direction calculated by the Bethe-Salpeter equation method. (c) Real and (d) imaginary part of the dielectric response function of monolayer h-BN in the out-of-plane direction by the Bethe-Salpeter equation method. Note that the dielectric responses of monolayer h-BN supercells with the thickness $d_{SC}$ of 20 Å and 40 Å are depicted in blue lines and pink areas, respectively.

## III. COMPUTATIONAL DETAILS

Calculations have been carried out with the Vienna ab initio simulation package (VASP) 5.4.4 code [48]. Structural optimization and ground-state Kohn-Sham energies are calculated using the plane-wave pseudopotential method based on the DFT with the Perdew-Burke-Ernzerhof (PBE) generalized gradient approximation (GGA) functional [49]. The long-range van der Waals interactions, which has a significant effect on the prediction of interlayer distance, are described via the Grimme-D3 scheme [50] with Becke-Johnson damping. The conjugate gradients algorithm is used in structural optimization with the maximum Hellmann-Feynman forces on each atom less than 0.001 eV/Å. Monkhorst-Pack grids of 8×8×6 and 12×12×1 are chosen for bulk and 2D systems, respectively. For 2D systems from monolayer to pentalayer, a sufficient vacuum layer larger than 20 Å is added between adjacent 2D layers to eliminate the spurious interlayer interactions under the periodic boundary condition. All values of thickness of 2D materials $d_{2D}$ and thickness of supercell $d_{SC}$ of h-BN and MoS$_2$ from monolayer to pentalayer can be found in Table 1. Single-shot $G_0W_0$ calculations are performed based on the screened Coulomb potential and polarizability constructed from the PBE-GGA wavefunctions and eigenvalues to obtain quasiparticle energies. The number of empty bands is set as 200 and 300 for bulk and 2D systems, respectively. The plane-wave cutoff energy is set as 400 eV for h-BN and 520 eV for MoS$_2$. Optical properties can be calculated based on the above wavefunctions and eigenvalues in the IPA picture [51].

As depicted in Fig. 3, h-BN and MoS$_2$ have hexagonal lattices with the in-plane lattice constants $a$ of 2.51 Å and 3.20 Å, respectively. We further consider the stable configurations for multilayer h-BN and MoS$_2$, which adopt A-A′ stacking order and trigonal prismatic (2H) configuration, respectively, with the genuine thickness $d_{2D}$ independent of vacuum layer shown in Figs. 3(c) and 3(f).

Optical excitation properties of h-BN and MoS$_2$ from monolayer to bulk are further calculated based on the MBPT [47] by solving the BSE [52,53] with the TDA [54,55]. The numbers of highest valence bands (m) and lowest conduction bands (*l*) to construct the transition space are selected as $m=$ 4, 8, 12, 16 and 8 for monolayer, bilayer, trilayer, tetralayer and bulk h-BN, respectively, and $m=$ 13, 26, 39, 52 and 26 for monolayer, bilayer, trilayer, tetralayer and bulk MoS$_2$, respectively; $l=$16 for bulk h-BN, 52 for bulk

MoS$_2$, and 60 for 2D systems. We have checked that our results are sufficiently converged with respect to these parameters for our discussion below. From the dielectric function evaluated by the IPA or BSE method, other optical properties including optical absorption α(ω), reflectivity R(ω), refraction index n(ω), extinction coefficient k(ω), and energy loss function L(ω) can be obtained through Eq. (11).

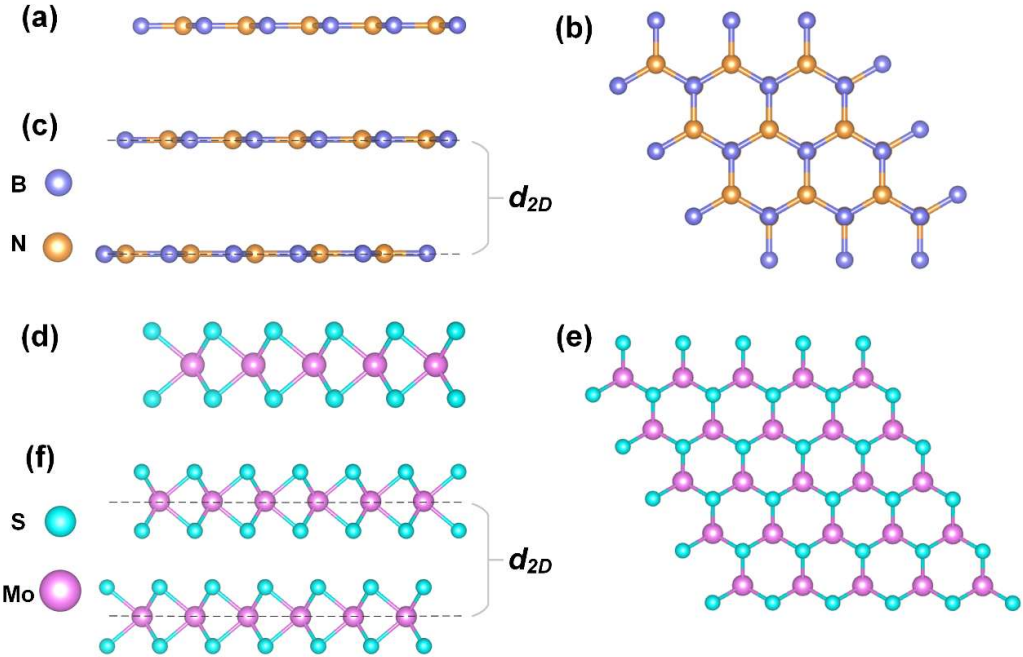

FIG. 3. (a) Side view and (b) top view of monolayer h-BN. (c) Bilayer h-BN stacked in A-A' order. (d) Side view and (e) top view of monolayer MoS$_2$. (f) Bilayer 2H-coordinated stack of MoS$_2$. Note that the genuine thickness of the single layer of 2D materials $d_{2D}$ is also denoted.

Table 1. Simulation parameters employed in the calculation of the dielectric and optical properties of h-BN and MoS$_2$, including the thicknesses of 2D materials and supercells as well as the out-of-plane lattice constants for bulk materials.

| Layer | h-BN | MoS$_2$ |
| --- | --- | --- |

|      | $d_{2D}$ (Å) | $d_{SC}$ (Å) | $d_{2D}$ (Å) | $d_{SC}$ (Å) |
|------|--------------|--------------|--------------|--------------|
| 1    | 3.39         | 23.85        | 6.32         | 32.04        |
| 2    | 6.65         | 27.71        | 12.18        | 42.25        |
| 3    | 9.93         | 30.03        | 18.06        | 49.74        |
| 4    | 13.95        | 41.56        | 24.05        | 60.83        |
| 5    | 16.51        | 45.41        | 30.09        | 66.92        |
| Bulk | 6.65         |              | 12.18        |              |

## IV. APPLICATIONS OF THE RESTORATION METHOD TO h-BN AND MoS$_2$

We first calculate the exciton binding energies of h-BN and MoS$_2$ from monolayer to bulk as the difference between the quasiparticle bandgaps and lowest excitation energies by the many-body GW-BSE method in Fig. 4. Monolayer h-BN demonstrates a large exciton binding energy of 2.08 eV and the exciton binding energy of few-layer h-BN decreases with the increasing thickness as shown in Fig. 4(a). Bulk h-BN is predicted to have a smaller exciton binding energy of 0.91 eV than the 2D counterparts, which is consistent with the nature of the relatively weakly-bound excitons in three-dimension bulk systems due to effective electronic screening. Likewise, monolayer MoS$_2$ demonstrates a large exciton binding energy of 0.73 eV and the exciton binding energies of few-layer MoS$_2$ show the same decreasing trend with the layer thickness as shown in Fig. 4(b). As for bulk MoS$_2$, it is predicted to have a smaller exciton binding energy of 0.05 eV.

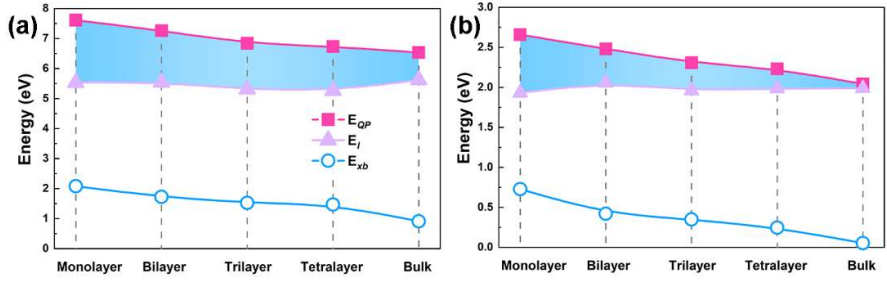

FIG. 4. The quasiparticle bandgaps $E_{QP}$, lowest excitation energies $E_l$, and exciton binding energies $E_{xb}$ of (a) h-BN and (b) $MoS_2$ from monolayer to bulk.

Intrinsic dielectric and optical properties of h-BN and $MoS_2$ from monolayer to bulk have been calculated by the BSE method in both in-plane and out-of-plane directions in Figs. 5 and 6. Compared with the results in the independent-particle picture, the BSE method yields different properties: the optical dielectric constants $\varepsilon_\infty$, static refraction indexes $n_0$, and static reflectivities $R_0$ calculated by the many-body method are higher in the in-plane direction and quite lower in the out-of-plane direction. Moreover, the above dielectric and optical properties of h-BN and $MoS_2$ show the same thickness dependence in the out-of-plane direction in Figs. 5(b) and 6(b). As for the in-plane direction, $\varepsilon_\infty$, $n_0$, and $R_0$ of h-BN and $MoS_2$ are found to change slightly with the layer thickness as shown in Figs. 5(a) and 6(a). It is worth noting that Laturia et al. [36] also calculated the intrinsic optical dielectric constants $\varepsilon_\infty$ of ultrathin h-BN and $MoS_2$. However, their restoration formula was based on the principle of equivalent capacitance, which is only suitable for calculating the static dielectric constant $\varepsilon_0$ (including both lattice contribution $\varepsilon_l$ and electronic contribution $\varepsilon_\infty$) and therefore cannot be applied to optical dielectric constant $\varepsilon_\infty$. According to the approach described in Sec. II, the in-plane and out-of-plane intrinsic optical dielectric constants $\varepsilon_\infty$ should be restored using Eq. (9), and the intrinsic static refraction indexes $n_0$ and static reflectivities $R_0$ can be obtained via Eq. (11).

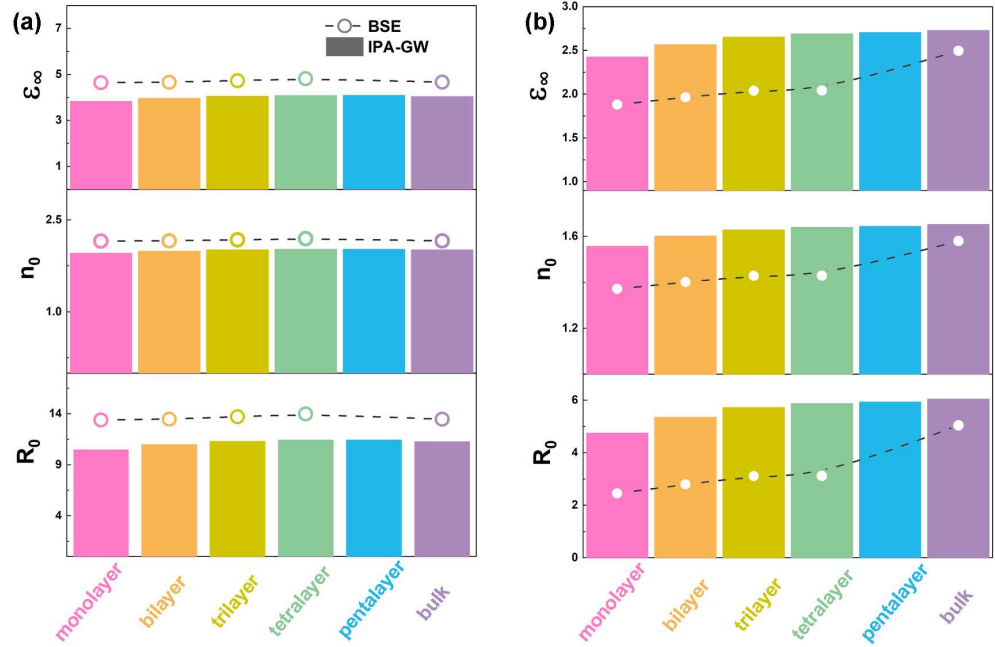

FIG. 5. (a) In-plane and (b) out-of-plane dielectric and optical properties of h-BN from monolayer to bulk including macroscopic optical dielectric constant static $\varepsilon_\infty$, static refraction indexes $n_0$, and static reflectivities $R_0$, calculated by the IPA-GW and BSE method depicted in bars and circles, respectively.

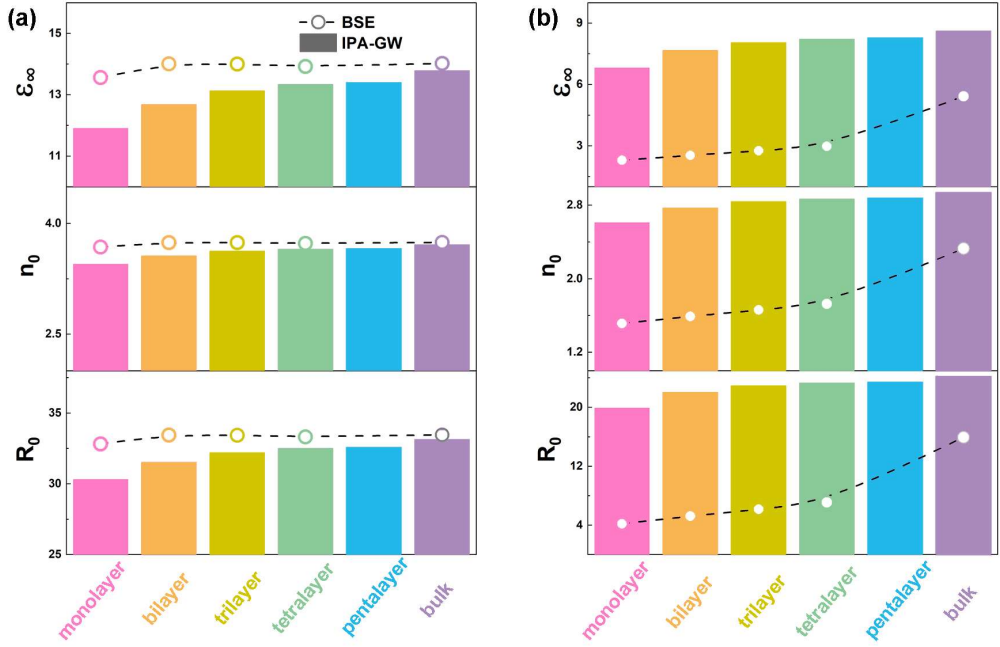

FIG. 6. (a) In-plane and (b) out-of-plane dielectric and optical properties of $MoS_2$ from monolayer to bulk including macroscopic optical dielectric constant static $\varepsilon_\infty$, static refraction indexes $n_0$, and static reflectivities $R_0$, calculated by the IPA-GW and BSE method depicted in bars and circles, respectively.

Furthermore, our method is also valid and effective to calculate the frequency-dependent dielectric functions of 2D materials. As examples, the intrinsic optical properties of monolayer, bilayer, and bulk h-BN and $MoS_2$ in in-plane and out-of-plane directions are shown as functions of photon energy in Figs. 7 and 8. Strong anisotropy can be found in the optical properties of both h-BN and $MoS_2$. In particular, in the out-of-plane direction, optical properties such as imaginary dielectric functions, reflectivities, extinction coefficients, and energy loss functions of 2D materials demonstrate lower values than the bulk counterparts.

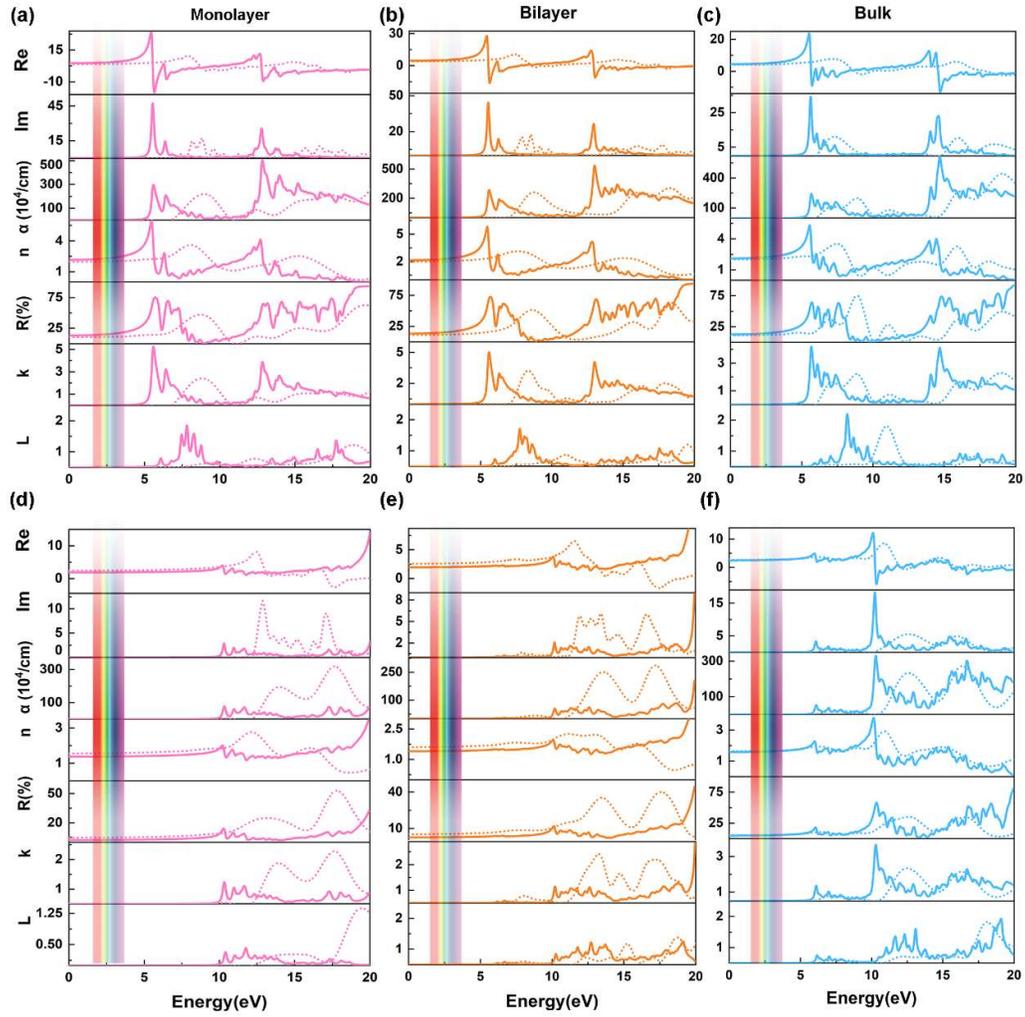

FIG. 7. In-plane excitonic optical properties including real and imaginary dielectric functions, optical absorption coefficients α(ω), refraction indexes n(ω), reflectivities R(ω), extinction coefficients k(ω) and energy loss functions L(ω) of h-BN from (a) monolayer, (b) bilayer to (c) bulk. Out-of-plane excitonic optical properties of h-BN from (d) monolayer, (e) bilayer to (f) bulk. Note that the dashed lines calculated in the independent-particle picture are shown for comparison.

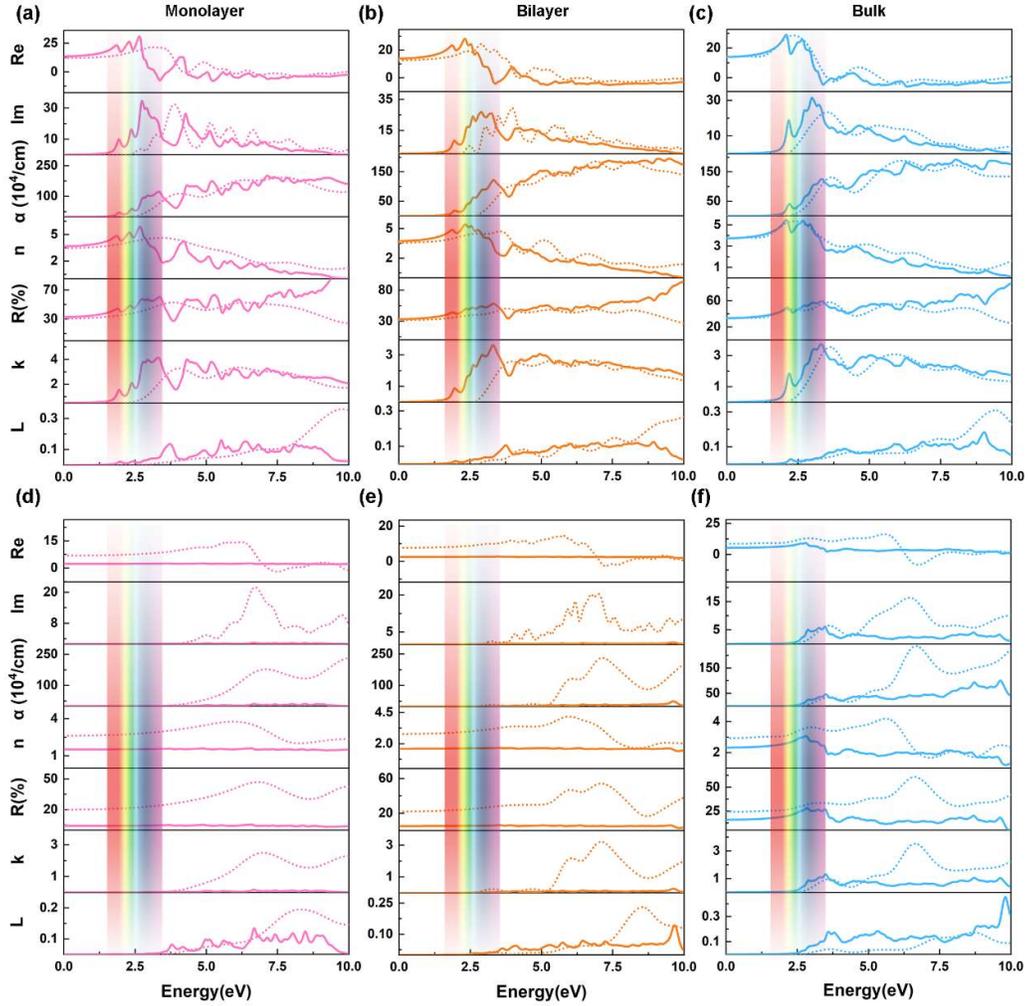

FIG. 8. In-plane excitonic optical properties including real and imaginary dielectric functions, optical absorption coefficients α(ω), refraction indexes n(ω), reflectivities R(ω), extinction coefficients k(ω) and energy loss functions L(ω) of $MoS_2$ from (a) monolayer, (b) bilayer to (c) bulk. Out-of-plane excitonic optical properties of $MoS_2$ from (d) monolayer, (e) bilayer to (f) bulk. Note that the dashed lines calculated in the independent-particle picture are shown for comparison.

For 2D materials, previous researches often calculated either the non-intrinsic optical properties [17,18,20,38–42] [such as the dotted lines in Fig. 9(a)], or only the relative shape of the imaginary dielectric functions to avoid the uncertainty introduced by the additional vacuum spacing [29,43–46]. In our work, however, we accurately predict the absolute value of the intrinsic optical properties of 2D h-BN and $MoS_2$ including dielectric functions, optical absorption coefficients, refraction indexes, reflectivities, extinction coefficients, and energy loss functions. More importantly, the calculated frequency-dependent properties could be compared with experimental measurements. As shown in Fig. 9, the predicted imaginary and real parts of the dielectric function, as well as the extinction coefficient and refraction index of monolayer $MoS_2$ all show great consistency with the experimental data [21] in green lines, boosting the validity and effectiveness of our restoration method to extensively and quantitively predict the dielectric and optical properties of other 2D materials.

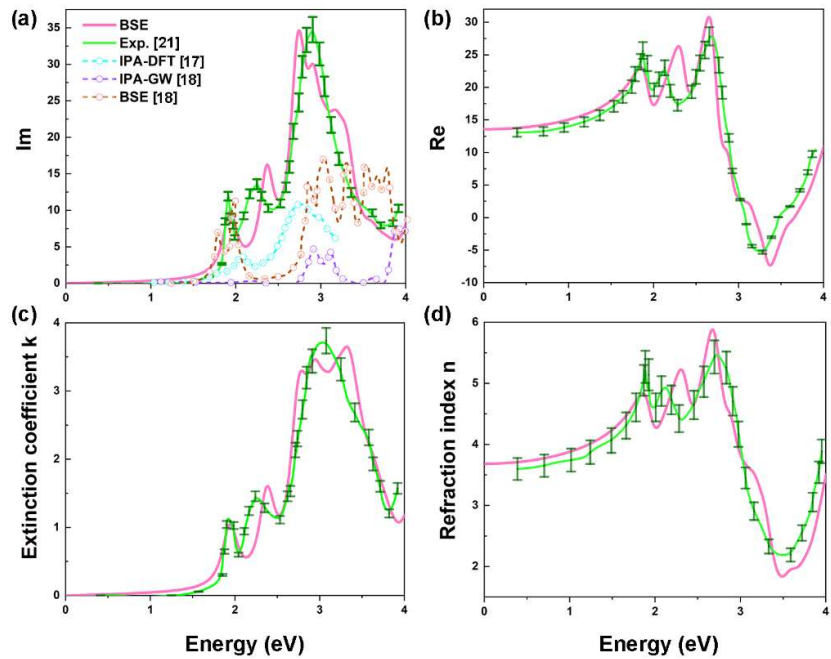

FIG. 9. In-plane (a) imaginary and (b) real parts of dielectric functions, (c) extinction coefficients, and (d) refraction indexes of monolayer $MoS_2$ from the BSE calculation in this work and experimental measurement [21]. Note that the error bars in the experimental curves are taken with an absolute value of ±5% due to different substrates used in the measurement. For the imaginary

dielectric function of in-plane monolayer MoS$_2$ in (a), previous non-intrinsic results calculated by the IPA-DFT, IPA-GW, and BSE methods [17,18] are illustrated for comparison.

## V. CONCLUSIONS

The influence of additional vacuum spacing in supercell calculations of the dielectric and optical properties of 2D materials has frequently been neglected and given rise to mispredictions far from the intrinsic properties of 2D materials or merely qualitative results. In this article, we have elucidated the relationship between the vacuum spacing and the dielectric functions of 2D materials in periodic calculations, which actually differs between the real and imaginary dielectric functions. Moreover, optical properties of 2D materials including optical absorption coefficients, refraction indexes, reflectivities, extinction coefficients, and energy loss functions based on the dielectric functions have demonstrated more complicated dependence on the vacuum spacing. As a result, we could restore the intrinsic dielectric and optical properties of 2D materials based on the relationship between the dielectric functions and the volumes of supercells we have put forward. Though simple in some sense, the restoration method is capable of correcting the popular errors in the previous non-intrinsic calculations of the dielectric and optical properties of 2D materials. As examples, we have calculated the intrinsic dielectric and optical properties of h-BN and MoS$_2$ from monolayer to pentalayer by the restoration method including dielectric functions, optical absorption coefficients, refraction indexes, reflectivities, extinction coefficients, and energy loss functions. Macroscopic optical dielectric constants, static refraction indexes, and static reflectivities of h-BN and MoS$_2$ have demonstrated thickness dependence in the out-of-plane direction in contrast to slight change with the thickness in the in-plane direction. We also find strong anisotropy in the dielectric and optical properties of h-BN and $MoS_2$, which differs between 2D and bulk counterparts. More importantly, our predictions demonstrate better consistency with measurements compared to previous calculations. Thereby, our restoration method shows great promise of extensively and quantitively predicting the intrinsic dielectric and optical properties of other 2D materials. Additionally, our method is also applicable to evaluate and compare previous non-intrinsic properties of 2D materials through explicit supercells volumes. We hope that this contribution could raise the importance of restoring intrinsic dielectric and optical properties of 2D materials and provide valuable insights into the calculations and investigations of various promising 2D materials.

## ACKNOWLEDGMENTS

This work was supported by the Natural Science Foundation of Shanghai (Grant No. 19ZR1404300). We thank An-An Sun for a careful reading of the text.

*gaosp@fudan.edu.cn